\documentclass{sig-alternate-05-2015}

\usepackage{amsmath, amssymb}

\newcommand\numberthis{\addtocounter{equation}{1}\tag{\theequation}}

\usepackage{hyperref}
\usepackage{xcolor}

\usepackage{booktabs}
\usepackage{tabularx}
\usepackage{hyperref}

\hypersetup{
    colorlinks,
    linkcolor={red!80!black},
    citecolor={blue!50!black},
    urlcolor={blue!80!black}
}

\def\sharedaffiliation{%
\end{tabular}
\begin{tabular}{c}}

\begin{document}


\doi{}

\isbn{}



%


\title{Exploring Model Predictive Control to Generate Optimal Control Policies for HRI Dynamical Systems}


 \numberofauthors{6} 
 \author{
 \alignauthor Steven Jens Jorgensen \\ 
 	   \affaddr{stevenjj@utexas.edu}
 \alignauthor Orion Campbell \\
 	   \affaddr{orioncampbell@utexas.edu}
 \alignauthor Travis Llado \\
 	   \affaddr{travisllado@utexas.edu}
 \and  
 \alignauthor Donghyun Kim\\
 	   \affaddr{dk6587@utexas.edu}
 \alignauthor Junhyeok Ahn\\
 	   \affaddr{junhyeokahn91@utexas.edu}
 \alignauthor Luis Sentis\\
 	   \affaddr{lsentis@austin.utexas.edu}       
 %
 \sharedaffiliation
        \affaddr{Department of Mechanical Engineering}\\
        \affaddr{The University of Texas at Austin}\\
        \affaddr{Austin, TX 78712}       
 }

\maketitle
\begin{abstract}
We model Human-Robot-Interaction (HRI) scenarios as linear dynamical systems and use Model Predictive Control (MPC) with mixed integer constraints to generate human-aware control policies. We motivate the approach by presenting two scenarios. The first involves an assistive robot that aims to maximize productivity while minimizing the human's workload, and the second involves a listening humanoid robot that manages its eye contact behavior to maximize ``connection'' and minimize social ``awkwardness'' with the human during the interaction. Our simulation results show that the robot generates useful behaviors as it finds control policies to minimize the specified cost function. Further, we implement the second scenario on a humanoid robot and test the eye contact scenario with 48 human participants to demonstrate and evaluate the desired controller behavior. The humanoid generated 25\% more eye contact when it was told to maximize connection over when it was told to maximize awkwardness. However, despite showing the desired behavior, there was no statistical difference between the participant's perceived connection with the humanoid.

\end{abstract}


\section{Introduction}
The study of HRI promises to create intelligent robots capable of becoming socially aware personal assistants. However, robots and assistive devices are not yet able to elicit the feeling of being understood in a human-like way. 

Our work is inspired from studying Social Cognitive Theory (SCT) \cite{Bandura1985} which claims that human behavior is based on the dynamic interplay of personal, environmental, and behavioral influences. It was recently used to model the walking exercise behavior of humans as a linear dynamical system \cite{Martin2014}. Among many other states that interplay with each other, their model included a measure of self-efficacy which increases as a result of exercise, thereby increasing the exercising behavior further. Subsequently, the authors also showed that a policy for behavior intervention can be generated using Model Predictive Control \cite{Martin2016}.

In the same way, we ask similar questions: Can HRI scenarios be modeled as dynamical systems? If so, can the tools of control theory generate useful policies? Previous work on modeling HRI scenarios and generating appropriate behaviors include creating belief models of the robot and human \cite{Breazeal2006, Gray2009}, probabilistic anticipatory action selection \cite{Hoffman2007}, collaborative agent planning \cite{Shah2011}, motion planning for navigation to maximize human comfort \cite{Sisbot2007}, fluent-turn taking using timed petri-nets \cite{Chao2016}, utilizing POMDPs for modeling cognition of an autonomous service robot \cite{Schmidt-Rohr2008}, and many others. For all scenarios the robot's cognitive model of the world and the human was necessary to generate appropriate actions to address the task at hand.

Here, we frame the cognitive modeling problem based on intuitive mechanical analogies. This technique leverages the power of feedback optimal control to generate useful interactive behaviors. In particular, this paper explores how Model Predictive Control (MPC) with mixed-integer constraints can be used to solve HRI scenarios modeled with linear dynamics. In doing so, this methodology is an approach towards creating \textit{cognitive feedback controllers}.


As the name suggests, MPC contains a model of the system and can simulate how its control policies can affect the model in the future \cite{Richards2005}. Since MPC can ``see'' a finite horizon into the future using its model of the world, it can identify locally what the best control policies are to minimize some objective function. 
In some HRI scenarios, a predictive controller provides significant advantages over traditional PID controllers.

\begin{figure}
\centerline{\includegraphics[width=0.8\columnwidth]{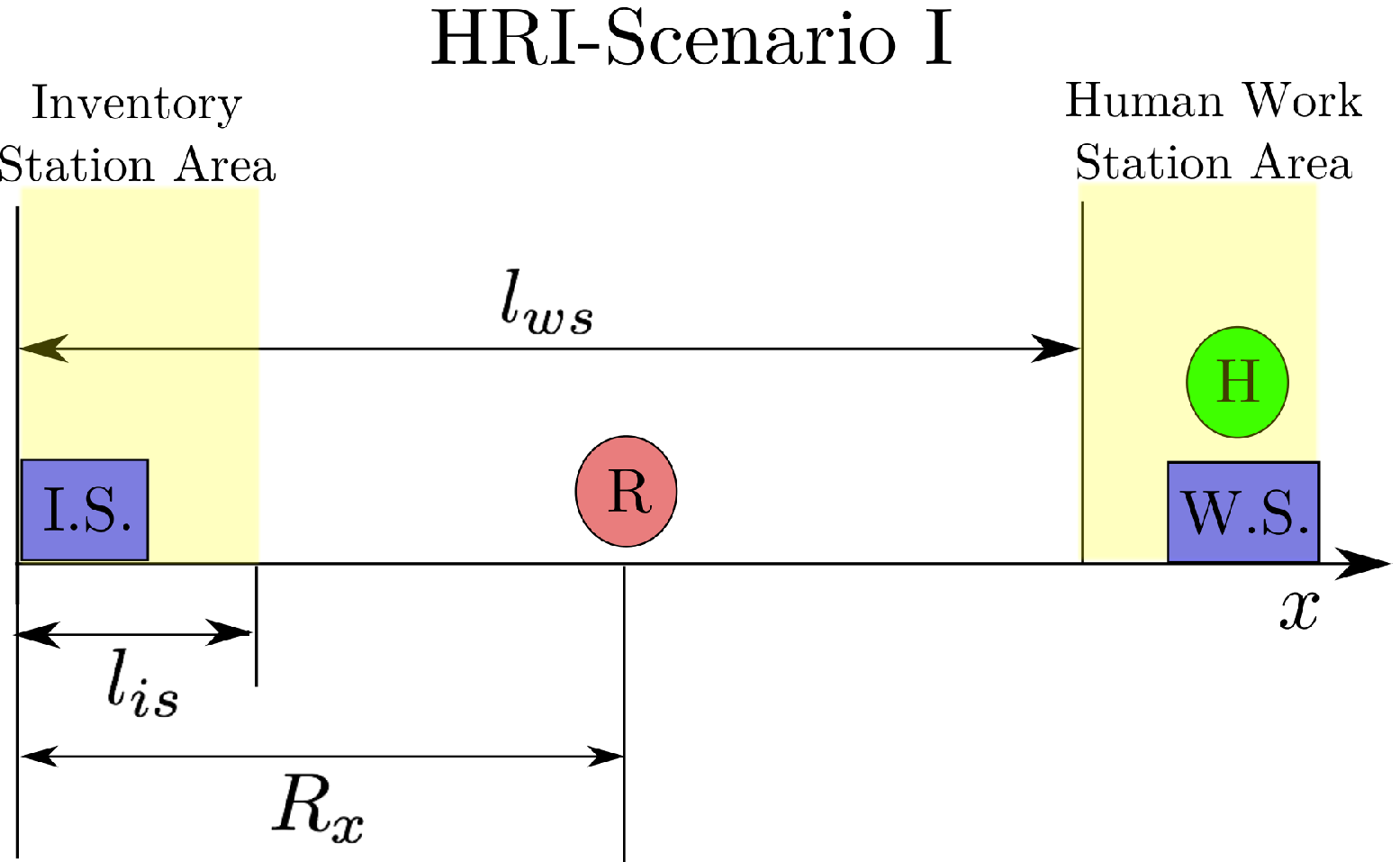}}
\caption{HRI Scenario I: \textmd{An assistive robot must bring deliverables from the inventory station (I.S.) to the human's work station area (W.S.). A mindful robot will ensure that the human is never overworked.}}
\label{fig:hri_scenarioI}
\end{figure}

\begin{figure}
\centerline{\includegraphics[width=1.0\columnwidth]{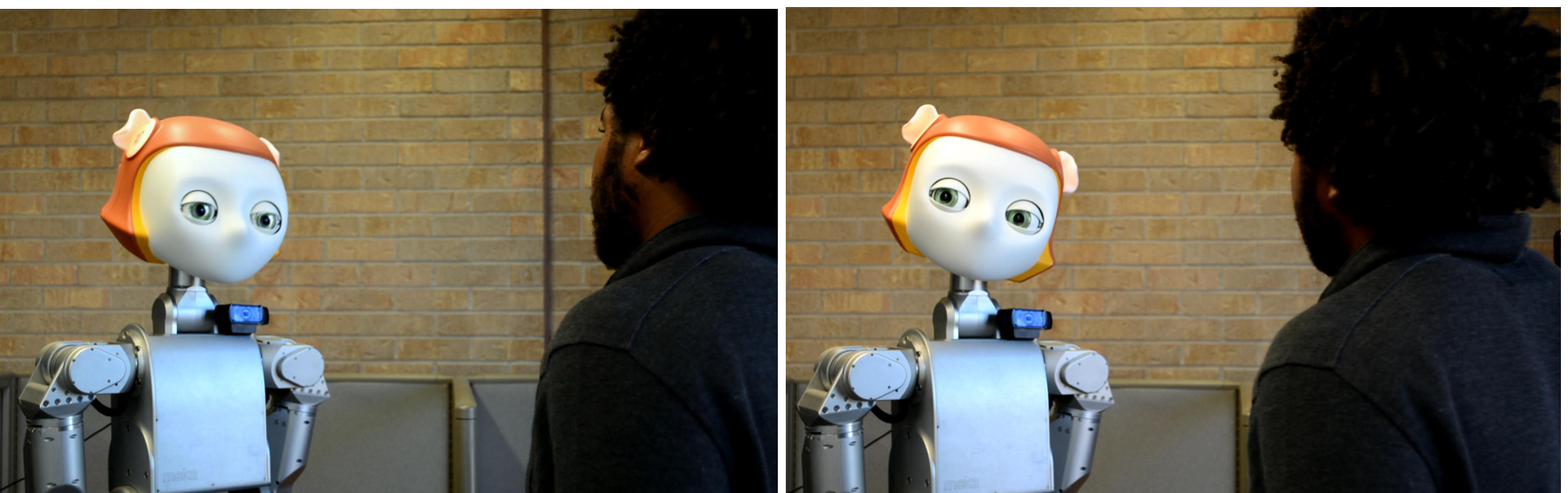}}
\caption{HRI Scenario II: \textmd{A humanoid listens to a participant's monologue about their weekly routine. Using the MPC controller, the humanoid manages its eye contact behavior to either maximize connection or awkwardness.}}
\label{fig:hri_scenario_II}
\end{figure}

To describe the usefulness of MPC for extracting useful control policies in HRI, we consider two scenarios. First, we consider an assistive robot that helps a human accomplish his work by bringing the human the necessary deliverables from an inventory station (Fig.~\ref{fig:hri_scenarioI}). The assistive robot must (1a) ensure it has enough battery to remain operational, (1b) continue being productive by delivering work to the human, and (1c) ensure that the human is never overworked. Second, we consider a listening humanoid robot which manages eye contact behavior (Fig.~\ref{fig:hri_scenario_II}) to (2a) maximize connection and (2b) minimize awkwardness with the human. 

Note that the two HRI scenarios contain a number of if-then statements which activate or deactivate boolean variables to specify if certain control inputs are available for the robot. For instance, the robot can only move \textit{if} it has enough battery power, or gaze contributes to awkwardness \textit{if} eye contact is maintained beyond a threshold value.  Such if-then constraints must be incorporated in the optimization routine mathematically via reformulating the statements as inequality constraints. To incorporate these constraints, we use Mixed Integer MPC, which is an optimization routine that minimizes some objective function subject to both real-valued and integer constraints. This optimization framework is called Mixed-Integer Programming (MIP) \cite{Smith2008}. 

Interestingly, while our objective function only contains battery levels, productivity, and human workload in the first scenario, and connection and awkwardness in the second, the robot is able to find the correct control policies to satisfy the end objective. 

To demonstrate the approach, simulations were performed using a Python interface called CVXPY  \cite{cvxpy}, a convex optimization library with mixed-integer programming capability with an academic license of the Gurobi \cite{gurobi} solver. Our code is available at \url{https://github.com/hrianon/mpc_hri}. Furthermore, the second scenario was implemented on a humanoid robot and tested with 48 human participants to demonstrate and evaluate the desired eye contact behavior.

\section{Related Works}
MIP was previously used for planning spacecraft trajectories \cite{Richards2002} and using integer constraints to model obstacle avoidance. MIP has also been used for generating optimal paths for manipulators \cite{Ding2011}.
Martin et al. \cite{Martin2014} used a fluid-tank analogy and a corresponding linear dynamic model to characterize human mental states that influence daily walking behavior. With a simplified version of the model, they controlled the system using Hybrid Model Predictive Control with integer and boolean constrains to achieve a desired goal \cite{Martin2016}.





In addition to previously mentioned works on modeling HRI and generating behaviors \cite{Breazeal2006, Gray2009, Hoffman2007, Shah2011, Sisbot2007, Chao2016, Schmidt-Rohr2008}, kinodynamic planning with RRT can also be used to solve search problems with dynamic constraints \cite{LaValle2001}. However, as with most planning algorithms, this requires specifying a desired goal state that may not be reachable. On the other hand, an MPC formulation performs an optimization routine to find the best control policy to minimize an objective function over the given time horizon. 

Proper eye contact duration, which can affect relationship and interpersonal affinity, was studied in \cite{Argyle1965}. Gaze type categorizations and computational understanding of social gaze was explored in \cite{Srinivasan2011}. The amount of precision and deliberate delays of the eye contact between robot and human were studied in \cite{Imai2002} and \cite{Admoni2014}. In addition, manipulating the eye contact frequency between a robot and human to make better connection was studied and implemented in \cite{Mutlu2006}. Here we propose a new model for controlling eye contact behavior using MPC. 

\section{Technical Approach}

\subsection{HRI as Linear Dynamical Systems}
To take advantage of the techniques found in the controls community, we model HRI scenarios as linear dynamical systems which have the form
\begin{equation}\label{eq:HRI_stateEqn_main}
\frac{dx}{dt} = \dot x = Ax + Bu,
\end{equation}
where $A \in \mathbb{A}^{n \times n}$ and $B \in \mathbb{R}^{n \times m}$describe the state changes due to the $n$ world states, $x \in \mathbb{R}^n $, and $m$ control inputs, $u \in \mathbb{R}^m$ respectively. In this paper, we discretize the dynamics by  $\Delta t$
\begin{equation} \label{eq:state_eq_main}
x(k+1) = (A\Delta t + I)x(k) + Bu(k)\Delta t,
\end{equation}
where $I \in \mathbb{R}^{n \times n}$ is identity and $x(k)$ and $u(k)$ denote the state and input at time step $k$.

\subsection{Policy Generation via MPC}
Given some desired robot behavior $y^{ref} \in \mathbb{R}^{n_y}$ with $n_y$ behaviors, e.g. we want the robot to be 100\% productive and minimize the human workload to 25\%, 
a standard quadratic cost function is used to quantify how well the decision vectors, $u_0, u_1, ..., u_{p-1}$ , bring a state output $y$ to $y_{ref}$ over a finite horizon $p$ time steps. The cost function is then defined as
\begin{equation}\label{eq:cost_function_main}
J = \sum_{i=1}^{p}  (y(k+i)-y^{ref})^T Q_y (y(k+i)-y^{ref}),
\end{equation}
where $Q_y \in \mathbb{R}^{n_y \times n_y}$ is a matrix describing the quadratic weights of the desired behavior.

The mixed-integer MPC problem can now be formulated as follows. In general, the MPC problem attempts to minimize a cost function $J$ subject to dynamic constraints and inequality constraints:

\begin{align*}
& \underset{ \{ [u(k+i)]_{i=0}^{p-1}, [\delta(k+i)]_{i=0}^{p-1} \} }{\text{argmin}} J \numberthis \label{eqn:argmin_scenario1} \\ 
& \text{s.t.} \\
& x(k+i) = (A\Delta t + I) x(k) + B u(k) \Delta t, \\
& y(k+i) = C x(k+i), \\
& E_1 \delta(k) \leq E_2 + E_3 x(k) + E_4u(k) + E_5 z(k),
\end{align*}



where $\delta(k) \in \{0,1\}^{n_{bool}}$ are $n_{bool}$ boolean variables used in the problem, $z(k) \in \mathbb{R}^{n_o}$ are $n_o$ floating variables, and  $E_{\{1,...,5\}}$, are matrices used to compactly specify the constraints of the problem. However, to be very clear about how constraints are specified in the MPC problem, we will describe each inequality constraint used for each scenario.

\section{Assistant Robot Scenario} \label{s:assist_robot_scenario}


We consider a scenario in which a robot assists a human with his work by delivering the necessary materials, tools, or paperwork needed by the human (deliverables) to perform useful work  (Fig.~\ref{fig:hri_scenarioI}). Subject to certain constraints described in Sec. \ref{sss:scenario1_constraints} The robot's primary role is to bring work to the human and its productivity is measured directly by how much work is delivered to the human. However, every time the robot drops off work to the human, the human's workload increases, so a mindful robot will be cautious of the human's workload.


\subsection{Linear Dynamic Model}
\subsubsection{World State and Actions}
We use a fluid-flow analogy to describe the linear dynamics of the scenario (Fig.~\ref{fig:tank_model_hri_scenarioI}) which enables easy visualization of how the different states of the world are affected by the robot's actions and other world states. We model the state of the world as a vector $x_R \in \mathbb{R}^5$,
\begin{equation}
x_R = [R_x, \ R_b, \ R_d, \ R_p, \ H_l]^T,  
\end{equation}
where $R_x$, $R_b$, $R_d$, $R_p$, and $H_l$, are the robot's x-coordinate position, battery levels, amount of deliverables being carried, self-perceived productivity, and perceived human workload respectively. In general, the system has control inputs $u \in \mathbb{R}^6$ defined as

\begin{equation}
u_R = [u_{move}, \ u_{charge}, \ u_{ipu}, \ u_{ido}, \ u_{wpu}, \ u_{wdo}]^T.  
\end{equation}

More specifically, $u_{move}$ lets the robot move left and right, $u_{charge}$ specifies how the robot's batteries change, $u_{ipu}$ and $u_{wpu}$ denote the act of picking up deliverables from the inventory station and the human's work station respectively, and $u_{ido}$ and $u_{wdo}$ denote the act of dropping off deliverables to the inventory station and the human's work station respectively.
 
\subsubsection{State Transition Matrix}
\begin{figure}
\centerline{\includegraphics[width=0.8\columnwidth]{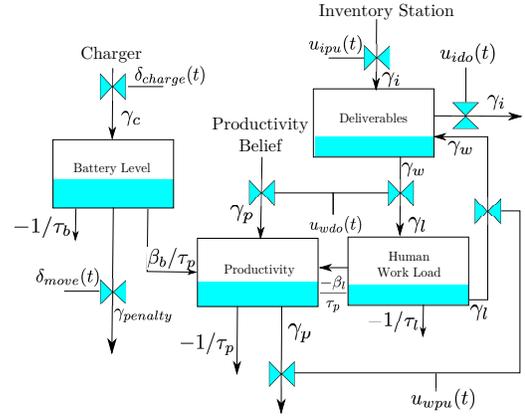}}
\caption{Fluid Analogy for Scenario I.}
\label{fig:tank_model_hri_scenarioI}
\end{figure}
Referring to Fig. (\ref{fig:tank_model_hri_scenarioI}) , at every time step, the robot's battery decreases by $-1/\tau_{b}$ and is further decreased as it moves ($|u_{move}| > 0$). The battery can be charged by $u_{charge}$ when the robot is near the inventory station. 

The robot feels productive whenever it has high battery levels, $\beta_B/\tau_p$, and whenever it drops off deliverables, $u_{wdo}$ to the human, but feels less productive whenever it takes work, $u_{wpo}$, away from the human and whenever the robot perceive high levels of human workload, $-\beta_l/\tau_p$. As with battery levels, the robot's perception of its own productivity decreases by $-1/\tau_{p}$ with time.

The robot's capacity to carry deliverables is modeled by the state $R_d$. Whenever the robot performs a deliverable pick up action ($u_{ipu}$ and $u_{wpu}$), this state increases. Similarly, when the robot performs a deliverable drop off action ($u_{ido}$ and $u_{wdo}$) this state decreases. 

Finally, the robot has a model of the human's workload, $H_l$. Namely, deliverables dropped off to the human's work station is analogous to increasing the human workload. The human, working at his own pace, reduces his workload by $-1/\tau_{l}$. 


Thus, the world state evolves with the following $A$ and $B$ matrices

\begin{equation}\label{eq:A_Matrix}
A = 
\left[ {\begin{array}{*{20}c}
   0 &     0        & 0 &   0       & 0  \\
   0 & -1/\tau_b    & 0 &   0       & 0  \\
   0 &     0        & 0 &   0       & 0  \\
   0 & \beta_b/\tau_p & -\beta_l/\tau_p & -1/\tau_p & 0  \\
   0 &     0        & 0 &   0       & -1/\tau_l  \\
 \end{array} } \right],
\end{equation}

\begin{equation}\label{eq:B_Matrix}
B = 
\left[ {\begin{array}{*{20}c}
   \gamma_{m}          &     0        & 0 &   0       & 0         & 0  \\
   0 & \gamma_{c}    & 0 &   0       & 0         & 0 \\
   0 &     0        & \gamma_{i} &   -\gamma_{i}       & -\gamma_{w}         &  -\gamma_{w} \\
   0 &     0        & 0            &    0                  & -\gamma_{p}       & \gamma_{p} \\
   0 &     0        & 0            &    0                  & -\gamma_{l}       & \gamma_{l} \\
 \end{array} } \right],
\end{equation}
where the $\gamma$ variables are constants that describe how the state changes due to the input $u$.

The current state transition matrix that describes the HRI scenario does not incorporate certain constraints of the problem. For example, the robot may only pick up deliverables whenever it is near the inventory station or when it is near the human work station. 
The next section describes how such constraints are automatically incorporated in the MPC problem statement. 


\subsection{MPC Formulation}
To maximize battery and productivity, and minimize human workload, we define the observation vector $y \in \mathbb{R}^3$ at time step $k+1$ to be $y(k+1) = Cx(k+1)$

where $C$ is
\begin{equation}\label{eq:C_Matrix}
C = 
\left[ {\begin{array}{*{20}c}
   0 &  1  & 0 &   0   & 0  \\
   0 &  0  & 0 &   1   & 0  \\
   0 &  0  & 0 &   0   & 1  \\
 \end{array} } \right],
\end{equation}

We set $y^{ref} = [1, 1, 0.25]^T$ which tells the robot to aim for 100\% battery level, 100\% productivity, and a human workload of 25\%. The cost function weights are set to be $Q_y = \text{diag} \{w_b, \ w_p, \ w_l\}$ with $w_b$, $w_p$, and $w_l$ denote the weights for the battery, robot productivity, and human workload respectively.

\subsubsection{Scenario I Constraints}\label{sss:scenario1_constraints}
The HRI scenario presented contains a number of constraints and the optimization routine must be restricted to a set of allowable actions depending on the world states. The robot has the following constraints:

\begin{enumerate}
  \item \label{const:S1_batterymove} It can move only if it has enough battery.
  \item \label{const:S1_batterycharge} Its batteries are charged only if it is near the inventory station.
  \item \label{const:S1_deliverableActions} It can  pick up and drop off deliverables only if it is near either station.
  \item \label{const:S1_capacity1} It cannot pick up deliverables beyond its capacity.
  \item \label{const:S1_capacity2} It cannot drop off deliverables if it has no deliverables. 
  \item \label{const:S1_movepenalty} It loses more battery as it moves.   
\end{enumerate}

Such if-then constraints must be converted to inequality constraints in order to frame the problem as a mixed-integer MPC problem. To specify constraint \ref{const:S1_batterymove}, we introduce a boolean variable, $\delta_{bat}$ that indicates if the robot has enough battery to move. Namely, it is only true if and only if the robot's battery is above a threshold, $b_{thresh}$.
\begin{equation} \label{eq:batteryLevel_ifthen}
\delta_{bat} = 1 \Leftrightarrow R_b(k) \geq b_{thresh}
\end{equation}
The following inequalities express this if-then constraint from Eq.~\ref{eq:batteryLevel_ifthen} as a mixed-integer constraint.
\begin{align}
R_b(k) - b_{thresh}  &\leq \delta_{bat}(k)(R_b^{max} - R_b^{min}), \\
R_b(k) - b_{thresh}  &\geq (1 - \delta_{bat}(k))(R_b^{min} - R_b^{max}),
\end{align}
where $R_b^{max}$ and $R_b^{min}$ are upper and lower bounds of the battery level. Next, we specify that the robot can only move if it has enough battery
\begin{align}
u_{move}(k) &\geq \delta_{bat}(k)u_{move}^{min}, \\
u_{move}(k) &\leq \delta_{bat}(k)u_{move}^{max},
\end{align}
where $u_{move}^{min}$ and $u_{move}^{max}$ specify the maximum movement effort the robot can use at every time step.

Next, to specify constraints \ref{const:S1_batterycharge}, and \ref{const:S1_deliverableActions}, we introduce two booleans, $\delta_{is}$ and $\delta_{ws}$ to indicate whether the robot is at the inventory station or the human work station. The desired location constraint is described as
\begin{align}
\delta_{is} &= 1 \Leftrightarrow R_x \leq l_{is} \\
\delta_{ws} &= 1 \Leftrightarrow R_x \geq l_{ws}
\end{align}
where $l_{is}$ specifies the location of the inventory station and $l_{ws}$ specifies the location of the human's work station. Similar to the battery level constraint, the location constraints can be expressed as a mixed-integer constraint using the following inequality constraints.
\begin{align}
(R_x(k) - l_{is}) &\leq (1-\delta_{is}(k)) R_x^{max}, \\
(R_x(k) - l_{is}) &\geq \delta_{is}(k) (-R_x^{max}), \\
(R_x(k) - l_{ws}) &\leq \delta_{ws}(k) R_x^{max}, \\
(R_x(k) - l_{ws}) &\geq (1-\delta_{ws}(k))  (-R_x^{max}).
\end{align}

Having location constraints, we can now constrain the robot to only pick up and drop off deliverables whenever it is near the inventory station or the human work station:
\begin{align}
0 \leq u_{ipu}(k) & \leq  \delta_{is}, \\
0 \leq u_{wpu}(k) & \leq  \delta_{ws}, \\
0 \leq u_{ido}(k) & \leq  \delta_{is}, \\
0 \leq u_{wdo}(k) & \leq  \delta_{ws}, 
\end{align}

Next, to specify capacity constraints \ref{const:S1_capacity1} and \ref{const:S1_capacity2}, we simply state that the robot cannot take actions that will cause it to exceed its carrying capacity of 100\% or to drop off deliverables when it doesn't have any.
\begin{align}
0 \leq R_d \leq 1.0
\end{align}

To specify constraint \ref{const:S1_movepenalty}, we introduce two boolean variables, $\delta_{mp}$ and $\delta_{mn}$ which are true if the robot exerts a positive effort and negative effort respectively. This is necessary because taking absolute values does not satisfy the Disciplined Convex Programming (DCP)~\cite{cvxpy} ruleset.
\begin{align}
\delta_{mp} = 1 \Leftrightarrow u_{move} > 0,
\delta_{mn} = 1 \Leftrightarrow u_{move} < 0.
\end{align}
These are again specified as inequality constraints
\begin{align}
(1 - \delta_{mp}(k))u_{move}^{min} < & u_{move}(k) <  \delta_{mp}(k)u_{move}^{max}, \\
\delta_{mn}(k)u_{move}^{min} < & u_{move}(k) < (1-\delta_{mn}(k))u_{move}^{max}.
\end{align}
Finally, to encode the overall change in battery due to robot movement and charging, we specify 
\begin{align}
u_{charge} = \delta_{charge} - (\delta_{move}) \gamma_{move penalty}
\end{align}
where $\delta_{charge} = \delta_{is}$, and $\delta_{move} = (\delta_{mp} + \delta_{mn})$. Additionally, to be consistent with the constants' effects as specified in Fig.~\ref{fig:tank_model_hri_scenarioI} and Eq.~\ref{eq:B_Matrix}, $\gamma_{move penalty} = \gamma_{penalty}/\gamma_{c}$. That is, the battery is charged whenever it is near the inventory station (constraint ~\ref{const:S1_batterycharge}) and the battery is decreased whenever the robot moves (constraint \ref{const:S1_movepenalty}).

\section{Eye Contact Scenario}
The second HRI scenario examined in this paper explores the dynamics of a one-on-one interaction between a human and a robot. 
The interaction characteristics we consider in this model are the \textit{connection} (or \textit{intimacy}) between the human and the robot and the \textit{awkwardness} of the exchange. 




\subsection{Linear Dynamic Model}
\begin{figure}
\centerline{\includegraphics[width=0.7\columnwidth]{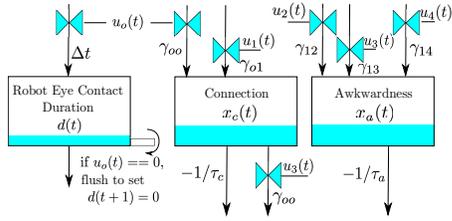}}
\caption{Fluid Analogy for Scenario II}
\label{fig:tank_model_hri_scenarioII}
\end{figure}
\subsubsection{World State and Actions}
Similar to Section \ref{s:assist_robot_scenario}, we use a fluid-flow analogy to describe the linear dynamics of the scenario (Fig.~\ref{fig:tank_model_hri_scenarioII}). The state of the interaction is a vector $x \in \mathbb{R}^2$,
\begin{equation}
x = [x_c, \ x_a]^T,  
\end{equation}

where $x_c$ is the ``human-robot connection'' or ``intimacy'' and $x_a$ is the ``awkwardness'' of the interaction. 

In this simple model, there are a variety of inputs that influence the state of the interaction, several of which can be either directly or indirectly controlled by the robot. The input $u \in \mathbb{B}^5$ is expressed as a boolean vector:
\begin{equation}
u = 
\left[ {\begin{array}{*{20}c}
   u_0 & = & \text{robot looking at person}  \\
   u_1 & = & \text{person looking at robot} \\
   u_2 & = & \text{gaze not reciprocated } (u_0 \neq u1) \\
   u_3 & = & \text{robot staring at person } (d > z) \\
   u_4 & = & \text{robot switching gaze } (u_0^{t} \neq u_0^{t-1}) \\ 
\end{array} } \right].  
\end{equation}

The auxiliary variable $d$ measures the duration of consecutive $u_0$ input activations (in other words, it tracks how long the robot has been looking at the person).  The staring threshold $z$ is a linear function of the connection state variable so that as connection increases, so does the amount of time that the robot can look at the human before the gaze would be classified as staring.  These variables change in discrete time according to the following equations:
\begin{equation} \label{eq:duration}
d(t+ \Delta t) = u_0(t)(d(t) + \Delta t),
\end{equation}
\begin{equation}\label{eq:staring_threshold}
z(t+ \Delta t) = m_z x_c(t) + b_z,
\end{equation}
where $\Delta t$ is the timestep and $m_z$ and $b_z$ are positive constants modulating the relationship between connection and the staring threshold such that if $d > z$, then the robot is considered to be staring at the person.

\subsubsection{State Transition Matrix}

Following the same method as the previous section, we describe how the state changes as a function of current states and inputs. At every time step, the connection and awkwardness decay by $-1/\tau_{c}$ and $-1/\tau_{a}$ respectively.

In each time step, gaze from the robot to the human and from the human to the robot contributes to the connection state by $\gamma_{00}$ and by $\gamma_{01}$ respectively.  Non-reciprocated gaze and staring contribute to awkwardness by $\gamma_{12}$ and $\gamma_{13}$ respectively. Additionally, robot staring cancels the contribution of the robot's gaze toward the connection. Finally, the switching gaze state detracts from or contributes to awkwardness (depending on the goal) by a factor of $\gamma_{14}$.  This last input is used to discourage a policy that would cause the robot to rapidly switch between looking at the person and looking away by slightly penalizing switching between the two states.

Thus, the world state evolves according to the same discretized state equation as the model in the previous section, except where the A matrix is

\begin{equation}\label{eq:A_Matrix_2}
A = 
\left[ {\begin{array}{*{20}c}
   -1/\tau_c    & 0           \\
   0            & -1/\tau_a   \\
 \end{array} } \right],
\end{equation}

and the B matrix is

\begin{equation}\label{eq:B_Matrix_2}
B = 
\left[ {\begin{array}{*{20}c}
   \gamma_{00}	&	\gamma_{01}	&	0			&	-\gamma_{00}	&	0			\\
   0 				&	0			&	\gamma_{12}	&	\gamma_{13}	&	\gamma_{14}	\\
 \end{array} } \right].
\end{equation}

\subsection{MPC Formulation}
With $y \in \mathbb{R}^2$ being our observation variable,  $C \in \mathbb{R}^{2 \times 2}$ an identity matrix relating $y$ to $x$ and $y^{ref}$ being the desired interaction state, we use the same standard quadratic cost function, $J$ in Eq.~\ref{eq:cost_function_main}, to quantify how well the input brings $y$ to $y^{ref}$ over a finite horizon $p$ time steps, and where $Q_y = \text{diag} \{w_c, \ w_a \}$ with $w_c$ and $w_a$ denote the cost function weights for the connection and awkwardness of the interaction.

We also assumed that during the prediction horizon $p$,  the human's gaze $u_1$ would not change.  Despite the fact that this assumption is probably false more often than not, we believe that if the control horizon is short enough, the policy will recover and can still produce usable behavior.


\subsubsection{Scenario II Constraints}

\begin{figure*}
\centerline{\includegraphics[width=15cm,height=15cm,keepaspectratio]{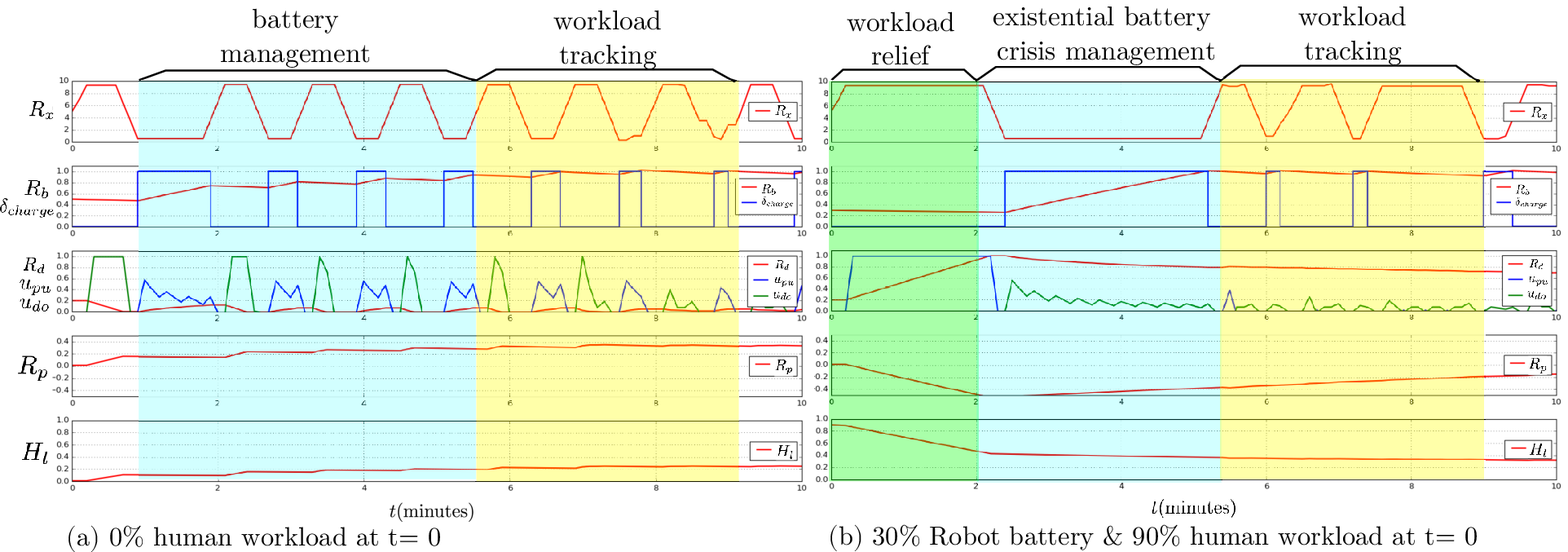}}
\caption{Assistant Robot Simulation Results: \textmd{For both (a) and (b), the robot worries more about the human's workload more than its own battery levels and productivity. Note that $u_{pu} = u_{ipu} + u_{wpu}$ and $u_{do} = u_{ido} + u_{wdo}$ to indicate the total deliverable pick up and drop off actions respectively. Also, the W.S. and  I.S. are located at $l_{ws} = 9$ and $l_{is} = 1$ respectively. In (a), the robot initially drops off the deliverables it is carrying to give the human work and proceeds to charge its own batteries while slowly dropping off more work to the human. In (b), the robot notices that the human is overworked and proceeds to remove work from the human at the cost of the robot's own productivity until the human's workload becomes manageable. The robot also charges its low battery levels to remain operational. Then, the robot proceeds to slowly drop off work to the human at a manageable rate, which also makes the robot's perception of its own productivity to rise again.} }
\label{fig:productivity_sim_results}
\end{figure*}

The constraints in this scenario ensure that the MPC formulation properly tracks the dynamics of the interaction as previously described.  The constraints are grouped as follows:
\begin{enumerate}
  \item \label{const:S2_state_equation} Enforce the discrete-time state equation.
  \item \label{const:S2_staring_threshold} Compute the staring gaze threshold according to equation \ref{eq:staring_threshold}.
  \item \label{const:S2_duration} Increment or reset the duration variable according to equation \ref{eq:duration}.
  \item \label{const:S2_human_gaze} Set the value of the human gaze input $u_1$.
  \item \label{const:S2_non_recip_input} Set the value of the non-reciprocated gaze input $u_2$.
  \item \label{const:S2_staring_input} Set the value of the staring gaze input $u_3$. 
  \item \label{const:S2_switching_input} Set the value of the switching gaze input $u_4$.
\end{enumerate}
\par
Constraints \ref{const:S2_state_equation} and \ref{const:S2_staring_threshold} are simply equality constraints enforcing equations \ref{eq:state_eq_main} and \ref{eq:staring_threshold}, respectively.

Constraint \ref{const:S2_duration} sets the tracking duration according to equation \ref{eq:duration}, which requires the following inequality constraints:
\begin{align}
u_0(k) d_{min} \leq &d(k+1) \leq u_0(k) d_{max} \\
(d(k)+\Delta t) - &d(k+1) \leq (1-u_0(k))(d_{max}-d_{min}) \\
(d(k)+\Delta t) - &d(k+1) \geq (1-u_0(k))(d_{min}-d_{max})
\end{align}
\par
Constraint \ref{const:S2_human_gaze} simply holds human gaze constant over the prediction horizon via
\begin{equation}\label{eq:human_eye_contact}
u_1(k) = u_1^{init}.
\end{equation}
Constraint \ref{const:S2_non_recip_input} can be expressed as an equality constraint
\begin{equation} \label{eq:equal_u2}
u_2(k) = u_0(k)(1-u_1^{init}) + (1-u_0(k))u_1^{init}.
\end{equation}
The multiplication of two variables normally would violate the rules of Disciplined Convex Programming (DCP); however, it is permissible in this case due to the fact that $u_1$ is held constant during the prediction horizon.  

In order to set the value of the staring gaze input $u_3$, two inequality constraints are necessary.
\begin{align}
	d(k) - z(k) &\leq      u_3(k)(d_{max} - z_{min}) \\
	d(k) - z(k) &\geq (1 - u_3(k))(d_{min} - z_{max})
\end{align}
\par 

The final set of constraints (\ref{const:S2_switching_input}) requires the introduction of two boolean auxiliary variables $\delta_0$ and $\delta_1$ which capture gaze turning on or off respectively. We then set $u_4$ by summing $\delta_0$ and $\delta_1$.
\begin{align}
u_4(k) = & \delta_0(k) + \delta_1(k) \\
& \delta_0(k+1) \leq u_0(k+1) \\
u_0(k+1) - u_0(k) \leq &\delta_0(k+1) \leq 1-u_0(k) \\
& \delta_1(k+1) \leq u_0(k) \\
u_0(k) - u_0(k+1) \leq &\delta_1(k+1) \leq 1-u_0(k+1)
\end{align}

\section{Simulation Results}

\begin{figure*}
\centerline{\includegraphics[width=15cm,height=15cm,keepaspectratio]{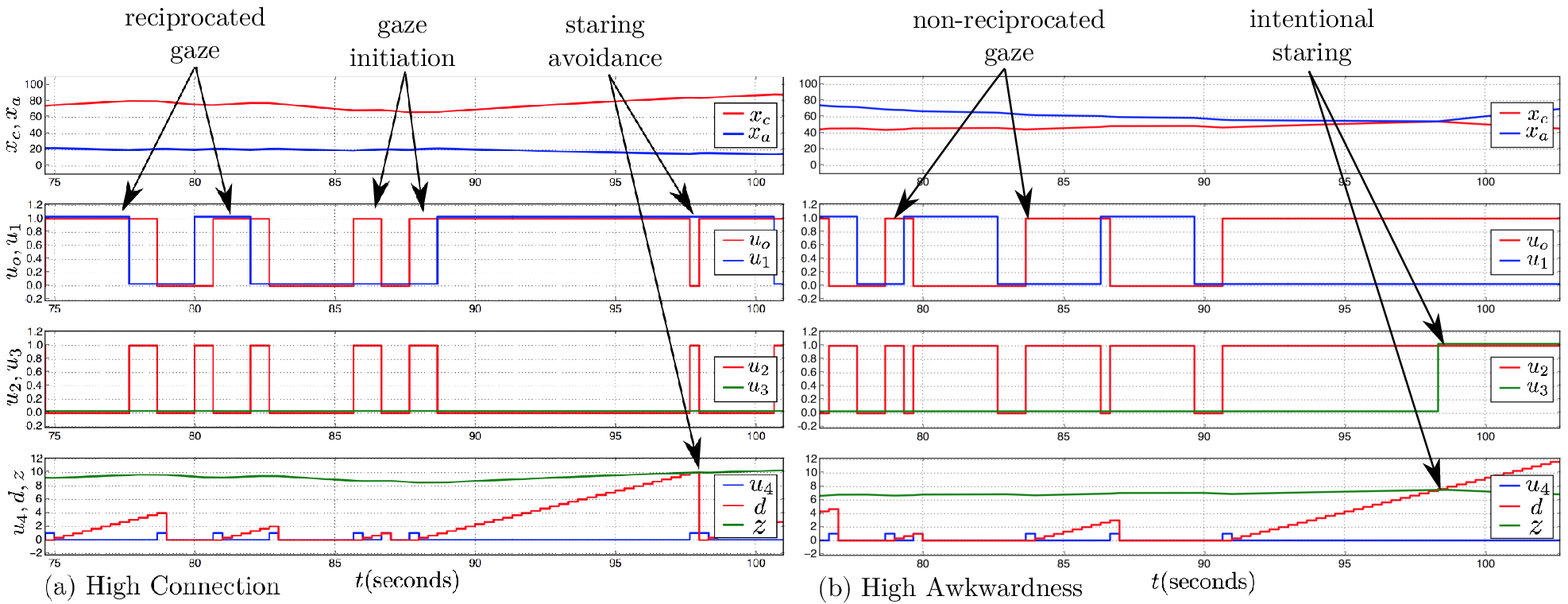}}
\caption{Eye Contact Behavior Simulation Results. \textmd{In (a), the robot is attempting to generate connection with the human while minimizing awkwardness. The robot does this by reciprocating gaze, and if the person looks away for an extended period, the robot will attempt to initiate gaze in order to bring the connection back up again, as shown between 85 and 90 seconds in this simulation. Finally, the robot will continue to hold gaze until just before it becomes awkwardly long as shown at ~97 seconds, at which point the robot glances away to release the tension held in mutual gaze. In (b) the robot is maximizing awkwardness via non-reciprocated gaze: the robot looks away when the person looks at it, and vice-versa.  Additionally, if given the chance, the robot will stare, as shown in the simulation plot beyond 97 seconds.} }
\label{fig:eye_contact_sim_results}
\end{figure*}

\subsection{Predictive Assistant Simulation}
We provide two test cases to the robot. For both cases, the optimization routine is set to maximize robot battery and productivity, and target a 25\% human workload ($y^{ref} = [1; 1; 0.25]$). The following weights $w_b$ = 1, $w_p =1$, and $w_l = 10$ were used. That is, the robot cares more about ensuring the human is never overworked over its own battery and productivity levels. The simulation parameters are available in the linked code repository.

In the first case (Fig.~\ref{fig:productivity_sim_results}a), the robot starts between the inventory station and the human work station and the human starts out with no workload. The robot first drops off its deliverables to the human and proceeds to charge its batteries. Then it moves back and forth to bring just enough deliverables to ensure that the human has a manageable workload (always at 25\%).

In the second case (Fig.~\ref{fig:productivity_sim_results}b), all the parameters and initial conditions are the same except that the human starts out with 90\% workload and the robot starts out with 30\% battery. Despite having low battery, the robot rushes to the human and removes the workload from the human. This causes the robot's productivity to become negative as per the definition of robot productivity. The robot understands that the human being overworked is the more important issue. When the human's workload is at 40\%, the robot charges its battery to remain operational. Then, the robot returns to a behavior which ensures the human has a manageable workload ($25-30\%$).%

\subsection{Eye Contact Simulation}

To test the eye contact model, we generated a random artificial human gaze pattern based on the geometric distribution such that the probability of the person switching his or her gaze in a single timestep was set to $\Delta t \cdot 20\%$ with $\Delta t = 0.33$ seconds.  We set the prediction horizon for the MPC to be 8 steps (2.66 seconds) and the control horizon (the number of control steps to be applied before recomputing the optimal control policy) to be 3 steps (1.0 second).

Two different interaction goal states were used for validation and tuning. In the first test case, the stated goal was to bring connection to 80\%, and to keep social awkwardness as close to 0\% as possible ($y^{ref} = [80,0]^T$). For this test case, both setpoints were given equal weight by setting $w_c = 1$ and $w_a = 1$ ($Q_y = \text{diag}\{1,1\}$).

For the second test case, the connection was ignored, and the awkwardness setpoint was increased to 80\%.  Thus, $y^{ref}= [0, 80]^T$ and $Q_y = \text{diag}\{0, 1\}$.
 
Fig \ref{fig:eye_contact_sim_results}(a) shows that when the robot aimed to create a connection, the robot would match the eye contact behavior of the human by reciprocating gaze. Additionally, the robot would initiate eye contact if the human had not looked for long periods of time in an attempt to build-up connection.

Fig \ref{fig:eye_contact_sim_results}(b) shows that when the robot aimed to maximize awkwardness, the robot generated non-reciprocated gaze behavior: the robot looked away when the human made eye contact and looked at the human when the human looks away. Additionally, the robot attempted to maximize awkwardness by staring at the person without disconnecting eye contact.

\section{Eye Contact Experiment}
To evaluate our eye contact MPC model, we test it on human participants. Our hypothesis is the following: ``If we aim for connection building behavior with our model, then participants will consider the interaction to be more natural than if we aim for awkward behaviors.''
\subsection{Human-Study Setup}
Participants were told to stand at a specified location in front of the humanoid and to describe their weekly routine for 60-90 seconds. Participants were only told that the humanoid is trying to learn how to interact better with humans. At the end of their monologue, participants filled out a short survey asking the level of enjoyment, comfort, and connection they felt with the robot as well as how much they believed the robot was interested in their story. For every participant, the robot's behavior was set to either maximize connection or maximize awkwardness.

While we implemented off-the-shelf face tracking, it was not reliable for detecting subtle eye contact changes. We therefore utilized the Wizard-of-Oz technique \cite{Riek2012} for wizard-recognition of eye contact through a small webcam mounted on the humanoid's shoulder. The operator would press the space bar key every time the human made eye contact with the robot. At a 3Hz control loop rate, this acted as the human eye contact, $u_1$, input to the MPC described in Eq.~\ref{eq:human_eye_contact}. 

The MPC outputs whether or not the robot should look at the human $u_0 \in \{0,1\}$. When $u_0 = 0$, the robot moves the head in a random direction outside a cone around the human's face, and otherwise the robot looks at the starting position to make ``eye contact'' with the human. Unfortunately, during in-lab testing, the humanoid's eyelid motors burned out, which may have affected our results.

\subsection{Results}
In the experiment, the robot produced the same types of behaviors described in the simulation section for each behavior target, except that it was in response to human input.

\begin{figure}
\centerline{\includegraphics[width=1.0\columnwidth]{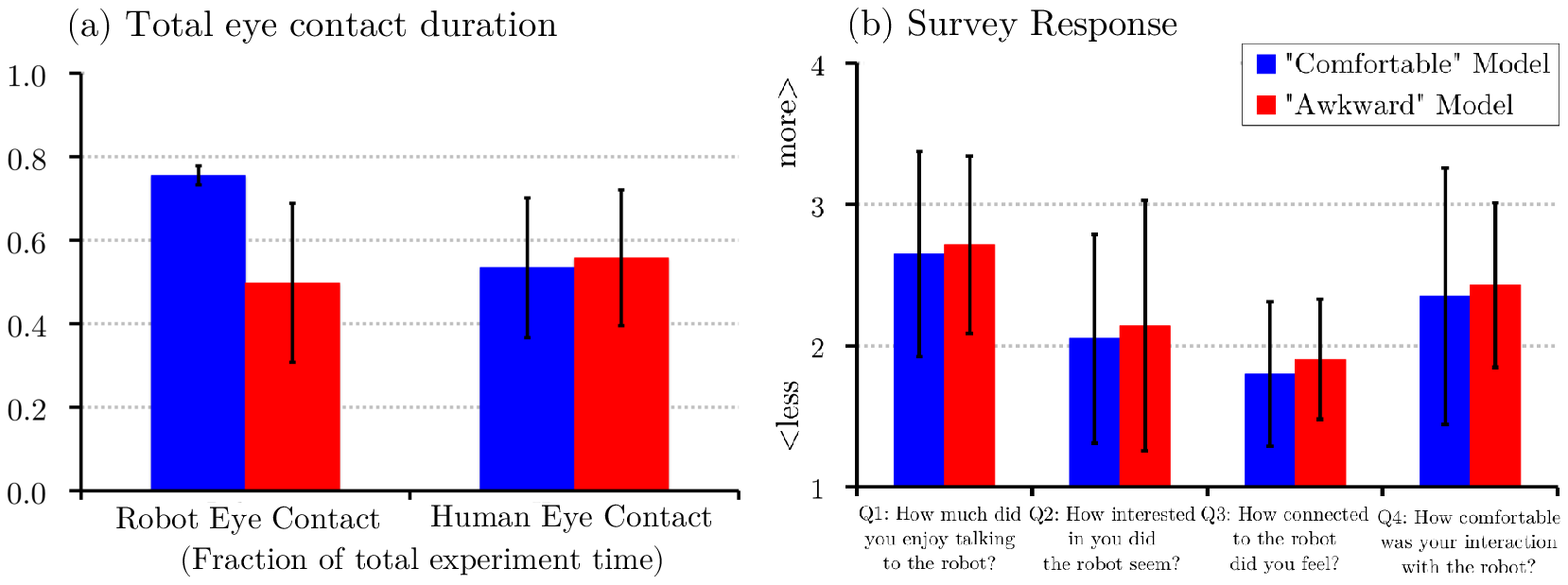}}
\caption{Experimental Results: \textmd{In (a), while the robot gave 25\% more eye contact to the participants, participants maintained similar eye contact behaviors with the robot. In (b), while no statistical significance is found, raw survey scores indicate that participants' responses were more favorable with an ``awkward'' robot. }}
\label{fig:exp_results}
\end{figure}

In terms of overall eye contact behavior, as Fig.~\ref{fig:exp_results}a shows,  the robot maintained eye contact $\sim$75\% of the overall interaction time when eliciting connection versus $\sim$50\% when attempting to maximize awkwardness. In terms of the participants perception, Fig.~\ref{fig:exp_results}b shows that there were no statistical differences between feeling more connection, comfort, enjoyment, or perceived interest when the robot behaved in either case. A correlation test for each survey question showed that the $R^2$ correlation values were $\{0.002, 0.003, 0.012, -0.003\}$ for questions q $\in \{Q1, Q2, Q3, Q4 \}$ (See Fig ~\ref{fig:exp_results}b). Due to these results, we must accept the null hypothesis that our MPC model did not produce significant effects on the participants for both robot behavior targets.

Interestingly, despite the similarity of the multiple choice responses, the 48 participants' written feedback were  more neutral in the connection eliciting behavior (4 positive, 7 neutral, 6 negative, and 7 non-responders) and more negative in the awkward eliciting behavior (2 positive, 6 neutral, 10 negative, and 6 non-responders).




\section{Discussions and Conclusions}
In the second scenario, while the robot demonstrated the desired behavior, participants did not feel more or less connection or awkwardness. This result could be due to a number of implementation factors such as the lack of eyelid control, the unmodeled effects of head pose, wizard error, the lag introduced by the control loop speed, and overly distracting look-away behaviors. Also, the model was tuned based on intuition. With these experimental results, human feedback could be used to better tune the model. Finally, many participants felt that the robot should be responding to their stories via backchanneling \cite{Rich2010} head movement, voice, and ear motions, which indicates that our eye contact model is not descriptive enough to elicit deep connection with humans. 

Potential future work includes testing the assistive robot MPC model as well as further improvements on modeling human-robot connection dynamics. Still, by taking inspiration from SCT models found in the exercise behavior intervention community, this paper explored the possibility of treating HRI as controllable dynamical systems in which state-of-the-art techniques from the controls community can be leveraged. Furthermore, the experiment showed that the control policies are deployable to real robotic systems.

\section{Acknowledgements}
This work was partially funded by a NASA Space Technology Research Fellowship, grant number NNX15AQ42H and an Office of Naval Research grant.

\bibliography{paper_bib}

\begin{thebibliography}{10}

\bibitem{Admoni2014}
H.~Admoni, A.~Dragan, S.~Srinivasa, and B.~Scassellati.
\newblock {Deliberate Delays During Robot-to-Human Handovers Improve Compliance
  With Gaze Communication}.
\newblock {\em International Conference on Human-Robot Interaction}, pages
  49--56, 2014.

\bibitem{Argyle1965}
M.~Argyle and J.~Dean.
\newblock {Eye-Contact, Distance and Affiliation}.
\newblock {\em Sociometry}, 28(3):284--304, 1965.

\bibitem{Bandura1985}
A.~Bandura.
\newblock {\em Social Foundations of Thought and Action: A Social Cognitive
  Theory}.
\newblock Prentice Hall, 1 edition, 1985.

\bibitem{Breazeal2006}
C.~Breazeal, M.~Berlin, A.~Brooks, J.~Gray, and A.~L. Thomaz.
\newblock {Using perspective taking to learn from ambiguous demonstrations}.
\newblock {\em Robotics and Autonomous Systems}, 54(5):385--393, 2006.

\bibitem{Gray2009}
C.~Breazeal, J.~Gray, and M.~Berlin.
\newblock {An Embodied Cognition Approach to Mindreading Skills for Socially
  Intelligent Robots}.
\newblock {\em The International Journal of Robotics Research}, 28(5):656--680,
  2009.

\bibitem{Chao2016}
C.~Chao and A.~Thomaz.
\newblock {Timed Petri nets for fluent turn-taking over multimodal interaction
  resources in human-robot collaboration}.
\newblock {\em The International Journal of Robotics Research}, page
  0278364915627291, 2016.

\bibitem{cvxpy}
S.~Diamond and S.~Boyd.
\newblock {CVXPY}: A {P}ython-embedded modeling language for convex
  optimization.
\newblock {\em Journal of Machine Learning Research}, 17(83):1--5, 2016.

\bibitem{Ding2011}
H.~Ding, G.~Rei{\ss}ig, D.~Gro{\ss}, and O.~Stursberg.
\newblock {Mixed-Integer Programming for Optimal Path Planning of Robotic
  Manipulators}.
\newblock {\em Int. Conference on Automation Science and Engineering}, pages
  133--138, 2011.

\bibitem{gurobi}
I.~Gurobi~Optimization.
\newblock Gurobi optimizer reference manual, 2015.

\bibitem{Hoffman2007}
G.~Hoffman and C.~Breazeal.
\newblock {Cost-Based Anticipatory Action Selection for Human-Robot Fluency}.
\newblock {\em IEEE Transactions on Robotics}, 23(5):952--961, 2007.

\bibitem{Imai2002}
M.~Imai, T.~Kanda, T.~Ono, H.~Ishiguro, and K.~Mase.
\newblock {Robot mediated round table: Analysis of the effect of robot's gaze}.
\newblock {\em Proceedings - IEEE International Workshop on Robot and Human
  Interactive Communication}, pages 411--416, 2002.

\bibitem{LaValle2001}
S.~M. LaValle and J.~J. Kuffner.
\newblock {Randomized Kinodynamic Planning}.
\newblock {\em International Journal of Robotics Research}, 20(5):378--400,
  2001.

\bibitem{Martin2016}
A.~Martin, D.~E. Rivera, and E.~B. Hekler.
\newblock {A Decision Framework for an Adaptive Behavioral Intervention for
  Physical Activity Using Hybrid Model Predictive Control}.
\newblock In {\em American Control Conference(ACC)}, pages 3576--3581, 2016.

\bibitem{Martin2014}
A.~Martin, D.~E. Rivera, W.~T. Riley, E.~B. Hekler, M.~P. Buman, M.~A. Adams,
  and A.~C. King.
\newblock {A Dynamical Systems Model of Social Cognitive Theory}.
\newblock In {\em American Control Conference(ACC)}, pages 2407--2412, 2014.

\bibitem{Mutlu2006}
B.~Mutlu, J.~Forlizzi, and J.~Hodgins.
\newblock {A Storytelling Robot : Modeling and Evaluation of Human-like Gaze
  Behavior}.
\newblock {\em 6th IEEE-RAS International Conference on Humanoid Robots}, pages
  518--523, 2006.

\bibitem{Rich2010}
C.~Rich, B.~Ponsler, A.~Holroyd, and C.~L. Sidner.
\newblock {Recognizing engagement in human-robot interaction}.
\newblock {\em Human-Robot Interaction (HRI), 2010 5th ACM/IEEE International
  Conference on}, pages 375--382, 2010.

\bibitem{Richards2005}
A.~Richards and J.~How.
\newblock {Mixed-integer programming for control}.
\newblock {\em Proceedings of the 2005, American Control Conference, 2005.},
  pages 2676--2683, 2005.

\bibitem{Richards2002}
A.~Richards, T.~Schouwenaars, J.~P. How, and E.~Feron.
\newblock {Spacecraft Trajectory Planning with Avoidance Constraints Using
  Mixed-Integer Linear Programming}.
\newblock {\em Journal of Guidance, Control, and Dynamics}, 25(4):755--764,
  2002.

\bibitem{Riek2012}
L.~D. Riek.
\newblock {Wizard of Oz Studies in HRI: A Systematic Review and New Reporting
  Guidelines}.
\newblock {\em Journal of Human-Robot Interaction}, 1(1):119--136, 2012.

\bibitem{Schmidt-Rohr2008}
S.~R. Schmidt-Rohr, M.~L{\"{o}}sch, and R.~Dillmann.
\newblock {Human and robot behavior modeling for probabilistic cognition of an
  autonomous service robot}.
\newblock {\em Proceedings of the 17th IEEE International Symposium on Robot
  and Human Interactive Communication, RO-MAN}, pages 635--640, 2008.

\bibitem{Shah2011}
J.~Shah, J.~Wiken, B.~Williams, and C.~Breazeal.
\newblock {Improved human-robot team performance using Chaski, a Human-Inspired
  Plan Execution System}.
\newblock {\em Proceedings of the 6th international conference on Human-robot
  interaction}, pages 29--36, 2011.

\bibitem{Sisbot2007}
E.~A. Sisbot, K.~F. Marin-Urias, R.~Alami, and T.~Sim{\'{e}}on.
\newblock {A human aware mobile robot motion planner}.
\newblock {\em IEEE Transactions on Robotics}, 23(5):874--883, 2007.

\bibitem{Smith2008}
J.~Smith and Z.~Taskin.
\newblock {A Tutorial Guide to Mixed Integer Programming Models and Solution
  Techniques}.
\newblock {\em Optimization in Medicine and Biology}, pages 1--23, 2008.

\bibitem{Srinivasan2011}
V.~Srinivasan and R.~R. Murphy.
\newblock {A survey of Social Gaze}.
\newblock {\em ACM/IEEE International Conference on Human-Robot Interaction},
  pages 253--254, 2011.

\end{thebibliography}
\bibliographystyle{abbrv}

\end{document}